\documentclass[runningheads]{llncs}
\usepackage{graphicx}
\usepackage{cite}
\usepackage{hhline}
\usepackage{booktabs}
\usepackage[numbers,sectionbib,square,comma]{natbib}
\usepackage{siunitx}
\usepackage{multirow}
\usepackage{enumitem}
\begin{document}
\title{Transfer Learning for Brain Tumor Segmentation}
%
\author{Jonas Wacker\inst{1} \and
Marcelo Ladeira\inst{2} \and Jose {Eduardo Vaz Nascimento}\inst{2,3}}
\authorrunning{J. Wacker et al.}
\institute{EURECOM, Biot, France\\
\email{jonas.wacker@gmail.com}\\(before with University of Brasília) \and
University of Brasília, Brasília-DF, Brazil \and
Syrian-Lebanese Hospital, Brasília-DF, Brazil}
\maketitle              
\begin{abstract}
Gliomas are the most common malignant brain tumors that are treated with chemoradiotherapy and surgery. Magnetic Resonance Imaging (MRI) is used by radiotherapists to manually segment brain lesions and to observe their development throughout the therapy. The manual image segmentation process is time-consuming and results tend to vary among different human raters. Therefore, there is a substantial demand for automatic image segmentation algorithms that produce a reliable and accurate segmentation of various brain tissue types. Recent advances in deep learning have led to convolutional neural network architectures that excel at various visual recognition tasks. They have been successfully applied to the medical context including medical image segmentation. In particular, fully convolutional networks (FCNs) such as the U-Net produce state-of-the-art results in the automatic segmentation of brain tumors. MRI brain scans are volumetric and exist in various co-registered modalities that serve as input channels for these FCN architectures. Training algorithms for brain tumor segmentation on this complex input requires large amounts of computational resources and is prone to overfitting. In this work, we construct FCNs with pretrained convolutional encoders. We show that we can stabilize the training process this way and achieve an improvement with respect to dice scores and Hausdorff distances. We also test our method on a privately obtained clinical dataset.

\end{abstract}
\section{Introduction}

According to the American Brain Tumor Association\footnote{\url{https://www.abta.org}}, in the United States alone, each year 68,470 people are diagnosed with a primary brain tumor and more than twice that number is diagnosed with a metastatic tumor. Gliomas belong to the group of primary brain tumors and represent around 28\% of all brain tumors. About 80\% of all the malignant (cancerous) tumors are gliomas, which makes them the most common malignant kind. They can be divided into low-grade (WHO grade II) and high-grade (WHO grades III-IV) gliomas \cite{Louis2016}, conforming to the World Health Organization (WHO) classification. 

Patients with the more aggressive high-grade gliomas (anaplastic astrocytomas and glioblastoma multiforme) have a median survival-rate of two years or less -- even with aggressive chemoradiotherapy and surgery.
The more slowly-growing low-grade (astrocytomas or oligodendrogliomas) variant comes with a life expectancy of several years \cite{Menze2015}.

In any case, Magnetic Resonance Imaging (MRI) modalities are used by radiotherapists before and during the treatment process. Brain tumor regions can be manually segmented into heterogeneous sub-regions (i.e., edema, enhancing and non-enhancing core) that become visible when comparing MRI modalities with different contrast levels \cite{Bakas2017, Bakas2017a, Bakas2017b}. Commonly employed MRI modalities include T1 (spin-lattice relaxation), T1c (contrast-enhanced), T2 (spin-spin relaxation), Fluid-Attenuated Inversion Recovery (FLAIR) and many others. Each modality  corresponds to gray-scale images that highlight different kinds of tissue.

The Brain Tumor Segmentation (BraTS) benchmark \cite{Menze2015, Bakas2018} revealed that there is a high disagreement among medical specialists when delineating the boundaries of various tumor subregions. Furthermore, the appearance of gliomas varies strongly among patients regarding shape, location and size.

Therefore, the segmentation task is considered quite challenging, especially when intensity gradients between lesions and healthy tissue are smooth. This is often the case since gliomas are infiltrative tumors.

Due to the high level of difficulty and the time-consuming nature of the task, there has been an emerging need for automatic segmentation methods over the last years. The goal of the BraTS benchmark is to compare these methods on a publicly available dataset. The data contains pre-operative multimodal MRI scans of high-grade (glioblastoma) and low-grade glioma patients acquired from different institutions.

\begin{figure}[t] 
\centerline{\includegraphics[width=0.8\linewidth]{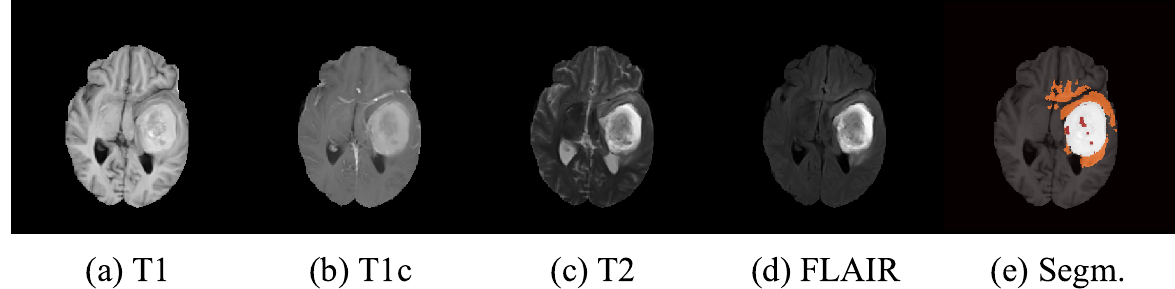}}
\caption{Example of a high-grade glioma case. The four MRI modalities (T1, T1c, T2, and FLAIR) of a single patient are shown (a-d). The rightmost image (e) shows the tumor segmentation. Orange corresponds to edema, white to the tumor core and red to the active tumor region.}
\label{fig:ex-segmentation}
\end{figure}

Fig.~\ref{fig:ex-segmentation} shows the four MRI modalities used in BraTS of an example patient along with the ground-truth annotations.

State-of-the-art methods for automatic brain tumor segmentation use fully-convolutional networks (FCNs) such as the U-Net \cite{Ronneberger2015} and its volumetric extensions (e.g. V-Net \cite{Milletari2016}). A major challenge in this regard is the volumetric multi-modal input that leads to high memory requirements and long training times despite the use of expensive GPUs.

In this work, we apply FCNs with encoders that are pretrained on the well-known ImageNet large-scale image dataset \cite{Russakovsky2015} in order to stabilize the training process and to improve prediction performance. This approach has led to outstanding results in two-dimensional image segmentation benchmarks such as the Carvana Image Masking Challenge\footnote{\url{https://www.kaggle.com/c/carvana-image-masking-challenge}}. The resulting U-Net architecture is called TernausNet \cite{Iglovikov2018}. We show that despite the difference between the ImageNet dataset and the MRI images used in this work, a performance gain can be achieved while stabilizing training convergence. The PyTorch implementation of our method is publicly available on GitHub\footnote{https://github.com/joneswack/brats-pretraining}.


In order to further evaluate our approach in a practical context, we have acquired MRI scans of five glioma patients from the Syrian-Lebanese hospital in Brasilia (Brazil) that we consider compatible with the BraTS data. We aligned this data with the scans of the BraTS challenge, which allows us to test our method on data that comes from an institution that did not provide any training or validation data.

\section{Method}
\label{sec:Methodology}

\subsection{Extending the AlbuNet architecture}
\begin{figure*}[h]
\centering
\includegraphics[width=1.0\textwidth]{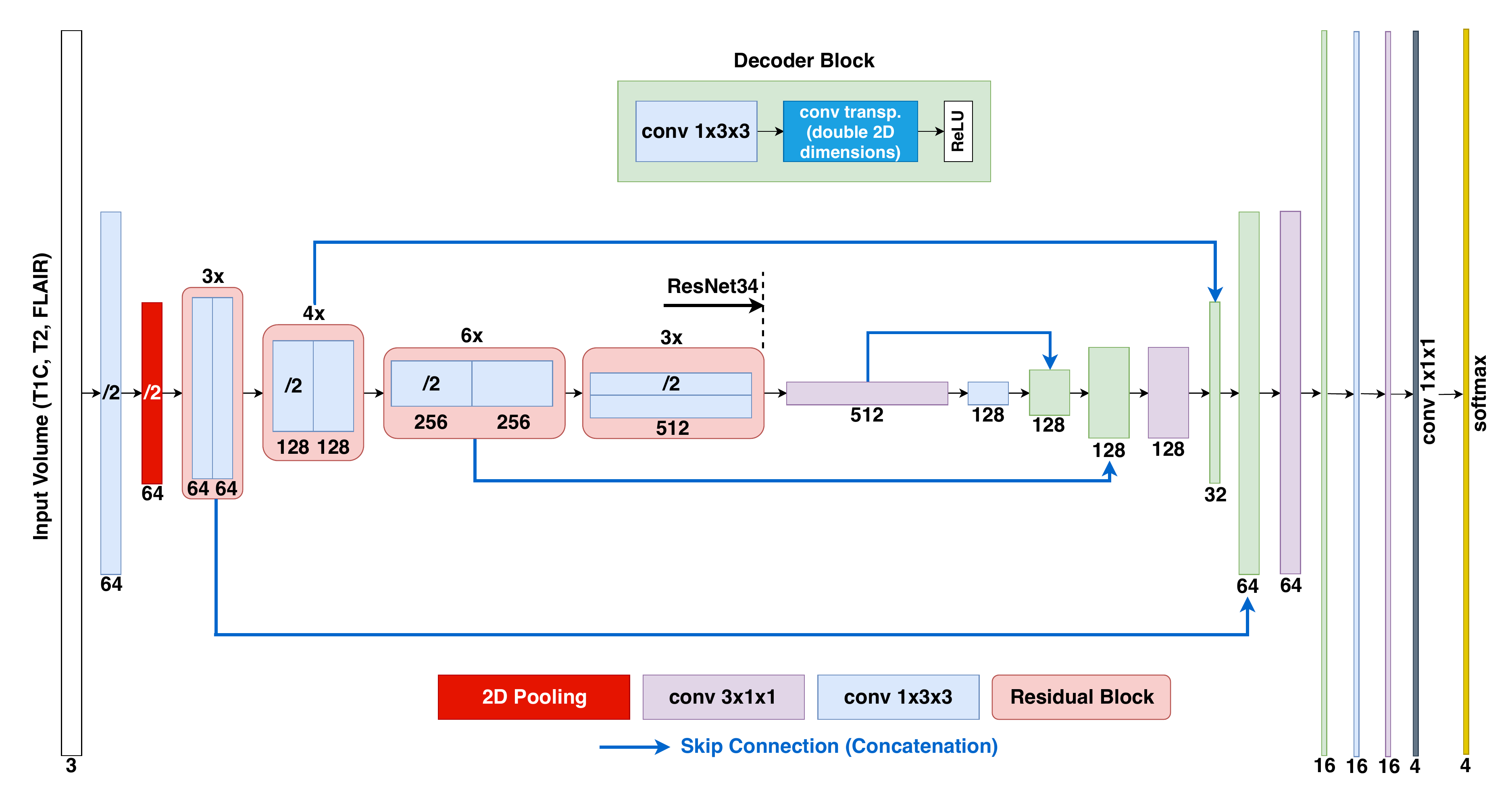}
\caption{The extended AlbuNet architecture. All 33 downsampling convolutional layers are extracted from the ResNet34 architecture that is pretrained on the ImageNet dataset. Until this point, the architecture is unmodified except for the initial pooling layer for which the kernel size is reduced from 3x3 to 2x2. Residual connections are maintained and only the final fully-connected layer is removed. The decoding blocks complement this downsampling path to retrieve equally dimensioned segmentation results. 3x1x1 depth layers are a novelty in our architecture.}
\label{fig:albunet}
\end{figure*}

We make use of a ResNet34~\cite{He2015} encoder that results in the AlbuNet variant~\cite{DBLP:journals/corr/abs-1804-08024} of TernausNet. Fig.~\ref{fig:albunet} depicts our U-Net based architecture. The ResNet34 downsampling layers are complemented with an upsampling path that uses transpose convolutional layers and receives intermediate inputs from the downsampling path. Both paths are symmetric in the dimensionality of intermediate outputs that they produce. However, no residual connections are used in the upsampling path that also contains less convolutional layers than the downsampling path.


In order to match the expected RGB three-channel input of the ResNet34 architecture, we select only the T1c, T2 and FLAIR modalities and discard T1 scans. We then normalize voxel intensities with respect to each scan. The resulting voxel intensities are directly treated as RGB channels, where R corresponds to FLAIR, G to T1c and B to T2.

We choose to discard the T1 modality without enhanced contrast instead of any other because this modality is rarely used by radiotherapists to delineate tumor boundaries. Furthermore, the privately acquired dataset presented in Section~\ref{sec:evaluation} does not contain T1 scans. However, including the T1 modality in our model may still improve predictions. We leave the extension of the architecture to 4-channel inputs for future research.

A novelty of our architecture is that we extend the 3x3 convolutions inside ResNet34 with 1x3x3 convolutions. Since the number of parameters stays the same, pretrained ResNet34 weights can still be loaded without modification. However, the effect is that the architecture is able to treat volumes of any depth now. The encoder simply processes the volumetric data slice-wise.

We add 3x1x1 convolutional depth layers in the upsampling path in order to exploit segmentation correlations in stacks of slices. The depth dimension can be changed to values different from 3 without further adaptation since the depth of the input is not reduced by this convolutional layer thanks to input padding. The architecture can be easily reduced to process slices instead of volumes by disabling the depth layers and choosing an input volume depth of one.

The last layer is a softmax over four channels where the channels correspond to the segmentation labels (edema, tumor core, enhancing tumor and background). Therefore, a single forward pass yields segmentations for all segmentation classes.

\subsection{Loss Function}
We use the Multiple Dice Loss as in \cite{Isensee2018} as a training objective:

\begin{equation}
\mathcal{L}_{DSC} = 1 - \frac{1}{K} \sum_{l=1}^{K} {DSC} \quad \text{with} \quad
DSC =  \frac{2\sum_{n} r_{ln} p_{ln}}{\sum_{n} r_{ln} + \sum_{n} p_{ln}},
\end{equation}

where $r_{ln}$ is the reference segmentation and $p_{ln}$ the predicted segmentation for voxel $n$ and class $l$. $DSC$ is the dice score coefficient that measures the similarity over two sets and can take on values between 0 (minimum) and 1 (maximum). Since there are three segmentation classes without background, we have $K=3$. The Multiple Dice Loss guarantees equal importance to each class irrespective of their proportion inside the scan. Using multiplications instead of set intersections and additions instead of set unions makes the dice loss function differentiable.


\subsection{Choice of Hyperparameters}
The batch size as well as the input patch size are determined by our hardware setup where we occupy the maximum amount of memory possible in order to speed up the training process and to exploit the trade-off between fast and well-directed weight updates.

For the case of 2D inputs, we use randomly sampled patches of 1x128x128 voxels, which leads to a batch size of 64. For the 3D case, we use a patch size of 24x128x128 voxels and a batch size of 24. The patch width and height are chosen to be roughly half of the input dimension of the data provided by the BraTS benchmark. However, it has to be noted that the brain of the patient only occupies a part of the space so that 128x128 patches cover a great part of it.

We use the Adam optimizer with a learning rate of $10^{-3}$ which yielded the best results in a preliminary evaluation. Training is carried out over 50 epochs where each epoch contains 100 input batch samples drawn uniformly at random from the non-zero area of the MRI input volumes.

\subsection{Preprocessing and Data Augmentation}
We normalize each voxel inside an input channel for each patient. Moreover, we crop the MRI input values to their non-zero regions in order to sample more efficiently and to reduce the memory footprint.

Finally, we apply a set of spatial transforms (data augmentation) to the input patches in order to improve the generalization error. These include elastic deformations, reflections and noise as well as blur. These transformations help the network to deal with unseen low-resolution scans inside the test dataset that occur frequently. The dataloader that we use for data agumentation and preprocessing is publicly available\footnote{https://github.com/MIC-DKFZ/batchgenerators}.

\subsection{Prediction}
We use the same patch sizes as the ones for training when predicting entire patient volumes. In order to improve prediction performance, we use a sliding window approach where predictions are carried out on overlapping input patches. The step sizes of the sliding window are 32 along the width and the height dimension and 24 along the depth dimension of the input. Overlapping predictions are averaged out yielding a slight gain in prediction performance compared to non-overlapping predictions.

\section{Evaluation}
\label{sec:evaluation}
\begin{figure}
\includegraphics[width=\textwidth]{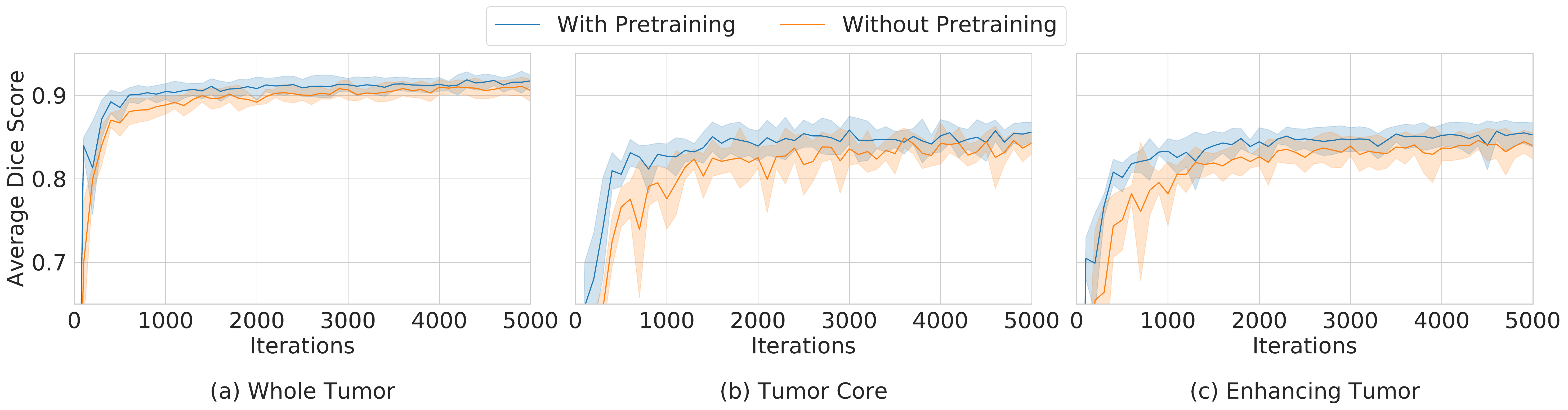}
\caption{Validation Dice Scores during training of the AlbuNet3D architecture. We carried out five training runs for the version with and without pretraining,respectively. Thick lines denote the mean dice score over five runs and the shaded error corresponds to the standard deviation from the mean.}
\label{fig:albu-convergence}
\end{figure}

We evaluate \textit{AlbuNet} (from now on denoted as \textit{AlbuNet2D}) as well as our proposed extension (\textit{AlbuNet3D}) with and without pretraining. Moreover, we compare these models on a privately acquired clinical dataset. For the private clinical dataset we only have access to the tumor core labels.




\subsection{Extended AlbuNet with and without pretraining}
We evaluate the effect of ImageNet pretraining on the brain tumor segmentation results for \textit{AlbuNet2D} and \textit{AlbuNet3D} as presented in Section~\ref{sec:Methodology}.

Fig.~\ref{fig:albu-convergence} shows validation dice scores during training of AlbuNet3D on the BraTS '20 training data with a held out validation set over 50 epochs. For each of the tumor regions the pretrained version keeps a small margin above the version without pretraining throughout the entire process. Initializing the model with pretrained weights also has a stabilizing effect on the training process as the standard deviations around the mean scores shrink. This holds in particular for the tumor core and enhancing tumor cases. Convergence happens more or less at the same speed and good results can be achieved after roughly 2000 iterations where each iteration processes one training batch of size 24 with patch size (24x128x128) voxels.


We used a single 16 GB GPU of the NVIDIA Tesla P100 graphics card for training. Training for 5000 iterations took 14 hours for \textit{AlbuNet3D} and 5.5 hours for \textit{AlbuNet2D} that follows a similar convergence pattern as the 3D architecture.

Using the prediction method presented in Section~\ref{sec:Methodology}, the prediction for a single patient takes 45 seconds for the \textit{AlbuNet2D} and 15 seconds for the \textit{AlbuNet3D} architecture respectively.

\subsection{Clinical dataset of the Syrian-Lebanese hospital}

We acquired a private dataset of 25 glioma patients from the Syrian-Lebanese hospital in Brasilia (Brazil). This dataset is not used for training and only serves evaluation purposes. The types of available MRI contrasts as well as their resolutions vary among patients. We selected 5 of these patients by hand for which T1c, T2 and FLAIR MRI modalities were provided. These were the most compatible ones with the BraTS data among the 25 patients. For none of the patients, the T1 modality used in the BraTS benchmark was available. Two of the five patients have two different T1c modalities, a volumetric one with a high depth resolution and a shallow one with a high axial resolution.

A great part of this work was to reverse-engineer the BraTS data preprocessing pipeline and to apply it to the 5 aforementioned patients.

This pipeline includes the following steps:
\begin{enumerate}[label=(\alph*)]
\item Conversion of MRI data as well as annotations from DICOM to NIFTI format using 3D Slicer\footnote{\url{https://www.slicer.org/}}.
\item{Conversion of each volume to an axial orientation using fslswapdim of the FMRIB Software Library (FSL) \cite{Jenkinson2012}}.
\item{Application of medical brain mask annotations to each scan in order to remove the skull of the patient.}
\item{Rigid Co-Registration of each brain volume to a reference scan that had the highest resolution (volumetric T1c in all cases) using nearest-neighbor interpolation. This was done using the FSL FLIRT linear co-registration tool~\cite{jenkinson2001}.}
\item{Resampling of all volumes and segmentations to an isotropic resolution (1mm x 1mm x 1mm) using nearest-neighbour interpolation.}
\end{enumerate}

For three of our five patients, only one T1c modality was available. This did not change the preprocessing in any way. The only difference is that there are only three modalities (volumetric T1c, T2 and FLAIR) instead of four (volumetric T1c, T1c, T2 and FLAIR). In the three cases where only one T1c modality was available we used this modality for both the T1 and the T1c input channel.


\subsection{Final Results}

\begin{figure}[t]
\centerline{\includegraphics[width=0.8\linewidth]{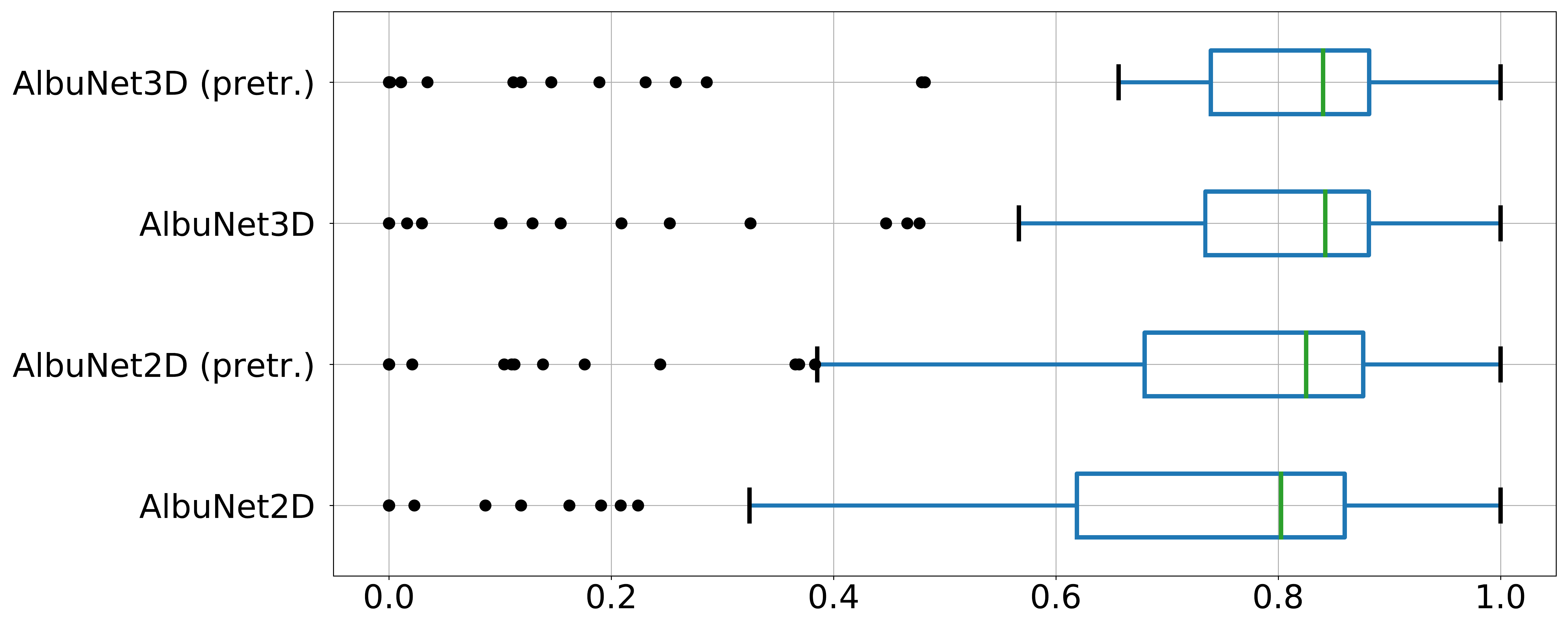}}
\caption{Boxplot of the enhancing tumor dice scores obtained from the evaluation on the BraTS'20 validation data.}
\label{fig:dice-boxplots}
\end{figure}

Fig.~\ref{fig:dice-boxplots} visualizes the \textit{enhancing tumor} dice score distribution for a single run of each method in a boxplot. It becomes apparent that our 3D extension of AlbuNet outperforms \textit{AlbuNet2D} (improved quartiles and medians). The use of pretrained weights has a positive effect for both the 2D and the 3D architectures. Nonetheless, there are a large number of outliers for each method that suggest a need for further improvements in robustness.

Table~\ref{tabl:validation_results} compares mean/median dice scores and Hausdorff distances for all segmentation labels and methods taken over 5 runs for the validation data and a single test run of the best-performing method. Except for the mean tumor core dice/Hausdorff scores, the pretrained AlbuNet3D architecture obtains the best validation results. The test score means are generally improved except for the whole tumor dice score while the median dice scores remain almost the same (except tumor core). Therefore, we can assume that the test set produced less outliers than the validation set. It also confirms that our method generalizes reasonably well to unseen data.

\begin{table}[htbp]
\caption{Mean (Median) dice scores and Hausdorff distances for the Brats'20 validation and test data.}
\label{tabl:validation_results}

\centering

\begin{tabular}{ccccccccccccc}

\toprule

\textbf{Method} &
\multicolumn{3}{c}{\textbf{Dice}} &
\multicolumn{3}{c}{\textbf{Hausdorff}} \\
& ET & WT & TC &
ET & WT & TC \\
\midrule

\multicolumn{7}{c}{\textbf{Validation Data (Average over 5 runs)}} \\
\midrule

\multirow{ 2}{*}{AlbuNet3D - pretrained} &
\textbf{0.7187} &
\textbf{0.8851} &
0.7822 &
\textbf{31.5322} &
\textbf{5.9929} &
15.6479 \\

& (\textbf{0.8382}) &
(\textbf{0.9127}) &
(\textbf{0.8830}) &
(\textbf{2.2361}) &
(\textbf{3.4308}) &
(\textbf{4.0196}) \\

\multirow{ 2}{*}{AlbuNet3D} &
0.7048 &
0.8808 &
\textbf{0.7926} &
39.0381 &
7.5833 &
\textbf{14.0151} \\

& (0.8321) &
(0.9079) &
(0.8775) &
(2.3214) &
(3.5406) &
(4.5917) \\

\multirow{ 2}{*}{AlbuNet2D - pretrained} &
0.6849 &
0.8786 &
0.7527 &
40.3850 &
7.9363 &
17.6747 \\

& (0.8222) &
(0.9043) &
(0.8700) &
(2.3972) &
(3.9187) &
(4.6775) \\

\multirow{ 2}{*}{AlbuNet2D} &
0.6842 &
0.8733 &
0.7447 &
38.6672 &
8.7802 &
19.4597 \\

& (0.8182) &
(0.9041) &
(0.8622) &
(2.4742) &
(3.8365) &
(5.1250) \\

\midrule
\multicolumn{7}{c}{\textbf{Test Data (Single final run)}} \\

\midrule

\multirow{ 2}{*}{AlbuNet3D - pretrained} &
0.7819 &
0.8744 &
0.8271 &
18.3230 &
5.4160 &
20.3297 \\

& (0.8373) &
(0.9117) &
(0.9158) &
(1.7321) &
(3.0000) &
(2.4495) \\

\bottomrule

\end{tabular}
\end{table}

Table~\ref{tabl:hospital_results} shows tumor core dice scores of five selected patients from our privately acquired clinical dataset whose scans were the most compatible with the BraTS challenge. The overall average performance (AVG) drops for all methods compared to the BraTS'20 evaluation in Table~\ref{tabl:validation_results}, which shows that it is not straightforward to apply existing methods to new clinical data.

The effect of pretraining is weaker on the tumor core labels as seen before. Since the clinical dataset is slightly different from the training data, these patients could be treated as outlier cases which is particularly true for patient 3 (lowest scores). A special treatment beyond pretraining is needed for outliers that is subject to further research.

\begin{table}[htbp]
\caption{Average tumor core dice scores and standard deviations over 5 runs for five patients (P1-P5) of the Syrian-Lebanese hospital.}
\label{tabl:hospital_results}
\centering
\begin{tabular}{>{\hspace{-2pt}}ccccccc<{\hspace{-2pt}}}
\toprule

\textbf{Method} &
\textbf{P1} &
\textbf{P2} &
\textbf{P3} &
\textbf{P4} &
\textbf{P5} &
\textbf{AVG}\\
\midrule
\multirow{ 2}{*}{AlbuNet3D - pretrained} & 0.8227 & 0.7918 & 0.4410 & 0.8112 & \textbf{0.7838} & 0.7301 \\
& ($\pm$0.06) & ($\pm$0.02) & ($\pm$0.10) & ($\pm$0.01) & ($\pm$0.04) & ($\pm$0.16) \\
\multirow{ 2}{*}{AlbuNet3D} & \textbf{0.8235} & \textbf{0.8046} & \textbf{0.4618} & \textbf{0.8190} & 0.7809 & \textbf{0.7380} \\
& ($\pm$0.02) & ($\pm$0.02) & ($\pm$0.09) & ($\pm$0.01) & ($\pm$ 0.03) & ($\pm0.16$) \\
\multirow{ 2}{*}{AlbuNet2D - pretrained} & 0.8042 & 0.7805 & 0.4026 & 0.8010 & 0.5855 & 0.6748 \\
& ($\pm$0.04) & ($\pm$0.02) & ($\pm$0.06) & ($\pm$0.01) & ($\pm$0.28) & ($\pm$0.18) \\
\multirow{ 2}{*}{AlbuNet2D} & 0.7939 & 0.7798 & 0.3631 & 0.8001 & 0.6217 & 0.6717 \\
& ($\pm$0.04) & ($\pm$0.03) & ($\pm$0.07) & ($\pm$0.02) & ($\pm$0.24) & ($\pm$0.19) \\
\bottomrule
\end{tabular}

\end{table}

\section{Conclusion}
\label{sec:conclusion}


We successfully extended \textit{AlbuNet2D} to process volumetric input patches and showed that \textit{AlbuNet3D} yields further performance improvements.

In order to outperform the state-of-the-art on the BraTS benchmark, future research should focus on the question of how to extend our method to all four input channels (T1, T1c, T2 and FLAIR). In this regard it would be possible to train an ensemble where each network uses a three-channel subset of the four provided channels. Another option would be to pretrain a model on four input channels, preferably on a large MRI dataset. This may improve the robustness of our method with respect to outliers.

We have shown that encoders pretrained on ImageNet improve the segmentation results of U-Net based architectures for the task of brain tumor segmentation and that they lead to a more robust training process. Unfortunately, this does not hold for our privately acquired clinical dataset where future research for robustness improvement is needed.
The available MRI data in a practical clinical context is much more heterogeneous than in the BraTS benchmark. Only a subset of the MRI modalities may be available for a patient and scans vary strongly in resolution, contrast and orientation. Therefore, it is crucial to continue our line of research to develop methods that can flexibly process changing input modalities.

\section{Acknowledgement}
We would like to thank the Syrian-Lebanese hospital in Brasilia for supplying us with MRI data for 25 glioma patients. We also appreciate the guidance that we received from their experts in the oncology department.

%
%
%
%




\bibliographystyle{splncs04}
\bibliography{brats_bib}
\end{document}